# A General Class of Regression Type Estimators in Systematic Sampling Under Non-Response

**Hemant Kr. Verma, R. D. Singh and †Rajesh Singh**

**Department of Statistics, Banaras Hindu University, Varanasi-221005, India**

**†Corresponding author**

**Abstract**

In this paper we have proposed a general class of modified regression type estimator in systematic sampling under non-response to estimate the population mean using auxiliary information. The expressions of bias and mean square error (MSE) up to the first order approximations are derived. A numerical study is included to support the theoretical results.

**1. Introduction**

In all sampling scheme, systematic sampling is most widely used due to its appealing simplicity. The method of systematic sampling first studied by Madow and Madow (1944) and is widely used in survey of finite populations. Estimation in systematic sampling has been discussed in detail by Lahiri (1954), Gautschi (1957), Hajeck (1959) and Cocharan (1957). Use of auxiliary information in construction of estimators is considered by Kushwaha and Singh (1989), Banarasi et.al. (1993), Singh and Singh (1998) and Singh et al. (2012).

Systematic sampling is a sampling technique in which first unit is selected randomly and the rest units are selected according to some predefined pattern. For example, suppose you want to take a sample of 8 houses from a street of 120 houses. 120/8=15, so every 15$^{th}$ house is chosen in your sample after a random starting point between 1 and 15. If the random start point is 13, then the houses selected are 13, 28, 43,58,73,88,103 and 118. (http://en.wikipedia.org/wiki/Systematic_sampling)

Let us suppose that a population consists of N units numbered from 1 to n in some order and a sample of size n is to be drawn such that N = nk (k is an integer). Thus there will be k samples each of n units and we select one sample from the set of k samples. Let Y and X be the study and auxiliary variable with respective means $\bar{Y}$ and $\bar{X}$. Let us consider $y_{ij}$ ($x_{ij}$) be the $j^{th}$ observation in the $i^{th}$ systematic sample under study (auxiliary) variable (i=1…k : j=1…n).

The usual regression estimator of the population mean $\bar{Y}$ based on a systematic sample of size n, under the assumption that the population mean $\bar{X}$ is known, is given by

$$\bar{y}_{lr}^* = \bar{y}^* + \beta\left(\bar{X} - \bar{x}^*\right) \tag{1.1}$$

where $\beta$ is the regression coefficient and is given by $\beta = \dfrac{s_{xy}}{s_x^2}$ and $\bar{y}^*$, $\bar{x}^*$ are estimators of population mean $\bar{Y}$ (study variable) and $\bar{X}$ (auxiliary variable), respectively, based on the systematic sample of size n.

The MSE of regression estimators $\bar{y}_{lr}^*$ is given by

$$MSE\left(\bar{y}_{lr}^*\right) = \theta \bar{Y}^2 \{1 + (n-1)\rho_X\}[C_Y^2 - K_1^2 C_X^2]\rho^{*2} \tag{1.2}$$

where, $\rho^* = \dfrac{\{1+(n-1)\rho_Y\}^{1/2}}{\{1+(n-1)\rho_X\}^{1/2}}$, $K_1 = \rho\dfrac{C_Y}{C_X}$, $\theta = \dfrac{1}{n} - \dfrac{1}{N}$

and $C_Y$, $C_X$ are the coefficients of variation of study of auxiliary variables respectively.

If some information is not available from sampling units or respondents due to various reasons in survey, for example, in an opinion survey, the family which is selected in our sample, might have shifted to some other place, selected person might have died. In mailed questionnaire, many respondents do not send their replies. So this type of problem is known as problem of non-response.

Since non-response is a serious problem in survey sampling. To deal with the problem of non-response, we divide the population in two groups, one is respondents and other is non-respondents. After that a sub sample is drawn from non-respondents group and then information is collected from this group. Two sample data are pooled to get the estimates of the population parameters ( Hansen and Hurwitz (1946).

Here, we assume that auxiliary variable is free from non-response and non-response is observed only on study variable. Using Hansen-Hurwitz (1946) technique of sub-sampling of non-respondents, the estimator of population mean $\bar{Y}$, can be defined as

$$\bar{y}^{**} = \frac{n_1 \bar{y}_{n1} + n_2 \bar{y}_{h_2}}{n} \tag{1.3}$$

where $\bar{y}_{n1}$ and $\bar{y}_{h_2}$ are, respectively the means based on $n_1$ respondent units from the systematic sample of n units and sub-sample of $h_2$ units selected from $n_2$ non-respondent units in the systematic sample. The estimator of population mean $\bar{X}$ of auxiliary variable based on the systematic sample of size n units, is given by

$$\bar{x}^* = \frac{1}{n}\sum_{j=1}^{n} x_{ij} \qquad (i = 1...k) \tag{1.4}$$

Obviously, $\bar{y}^{**}$ and $\bar{x}^*$ are unbiased estimators. The variance expression for the estimators $\bar{y}^{**}$ and $\bar{x}^*$ are, respectively, given by

$$V\left(\bar{y}^{**}\right) = \theta(1 + (n-1)\rho_Y)S_Y^2 + \frac{L-1}{n}KS_{Y2}^2 \tag{1.5}$$

$$V\left(\bar{x}^*\right) = \theta(1 + (n-1)\rho_x)S_x^2 \tag{1.6}$$

Where $\rho_Y$ and $\rho_x$ are the correlation coefficients between a pair of units within the systematic sample for the study and auxiliary variables respectively. $S_Y^2$ and $S_x^2$ are respectively, the mean square of the entire group for study and auxiliary variable. $S_{Y2}^2$ be the mean square of non-response group under study variable, K is the non-response rate in the population and $L = \dfrac{n_2}{h_2}$.

The regression estimators defined in equation (1.1) under non-response can be written as

$$\bar{y}_{lr}^{**} = \bar{y}^{**} + \beta\left(\bar{X} - \bar{x}^*\right) \qquad (1.7)$$

The MSE expression for the above estimator is given by

$$MSE\left(\bar{y}_{lr}^{**}\right) = \theta \bar{Y}^2 \left\{[1 + (n-1)\rho_x]C_Y^2 - K_1^2 C_X^2\right\}^{*2} + \dfrac{L-1}{n}W_2 S_{Y2}^2 \qquad (1.8)$$

In this paper we have proposed a general class of modified regression type estimator for estimating the population mean in systematic sampling using auxiliary information in the presence of non-response. A numerical study is carried out to compare the optimum estimator with respect to usual mean estimator with the help of numerical problem.

## 2. Proposed class of estimators:

Adapting Rao (1991) estimator, here we propose an estimator $t_1$ as

$$t_1 = w_{11}\bar{y}^{**} + w_{12}(\bar{X} - \bar{x}^*) \qquad (2.1)$$

Where $w_{11}$ and $w_{12}$ are constants.

An estimator $t_2$ can be defined as

$$t_2 = \left[w_{21}\bar{y}^{**} + w_{22}(\bar{X} - \bar{x}^*)\right]\left[\dfrac{\bar{X}}{\alpha\bar{X} + (1-\alpha)\bar{x}^*}\right]^{\delta} \qquad (2.2)$$

where $\alpha, \delta, w_{21}$ and $w_{22}$ are constants.

Adapting Bedi and Hajela (1984) estimator, we propose an estimator $t_3$ as

$$t_3 = \gamma\left[\bar{y}^{**} + b(\bar{X} - \bar{x}^*)\right] \qquad (2.3)$$

where b is regression coefficient and $\gamma$ is a constant.

To obtain the expression of bias and MSE of the estimators $t_1$, $t_2$ and $t_3$, let

$$\bar{y}^{**} = \bar{Y}(1+e_0)$$

$$\bar{x}^{*} = \bar{X}(1+e_1)$$

such that $E(e_0) = E(e_1) = 0$

and

$$E(e_0^2) = \frac{V(\bar{y}^{**})}{\bar{Y}^2} = \theta\{1+(n-1)\rho_Y\}C_Y^2 + \frac{L-1}{n}K\frac{S_{Y2}^2}{\bar{Y}^2},$$

$$E(e_1^2) = \frac{V(\bar{x}^{*})}{\bar{X}^2} = \theta\{1+(n-1)\rho_X\}C_X^2,$$

and $E(e_0 e_1) = \theta\{1+(n-1)\rho_Y\}^{½}\{1+(n-1)\rho_X\}^{½}\rho C_Y C_X$.

The bias expression for the estimators $t_1$, $t_2$ and $t_3$ are respectively, given by

$$\text{Bias}(t_1) = (w_{11} - 1)\bar{Y} \tag{2.4}$$

$$\text{Bias}(t_2) = \bar{Y}\left[w_{21}\left\{1 - \theta\delta\rho C_0 C_1 + \frac{\delta(\delta+1)}{2}\theta C_1^2\right\} - 1\right] + \delta w_{22}\bar{X}\theta C_1^2 \tag{2.5}$$

$$\text{Bias}(t_3) = (\gamma - 1)\bar{Y} \tag{2.6}$$

Similarly, the expressions of MSE's of the above estimators are given by

$$\text{MSE}(t_1) = \bar{Y}^2 + w_{11}^2 \bar{Y}^2(1+\theta C_0^2) + w_{12}^2 \bar{X}^2 \theta C_1^2 - 2w_{11}w_{12}\bar{X}\bar{Y}\theta\rho C_0 C_1 - 2w_{11}\bar{Y}^2$$
$$+ \frac{L-1}{n}KS_{Y2}^2 w_{11}^2 \tag{2.7}$$

Partially differentiating the above expression with respect to $w_{11}$ and $w_{12}$, and then equating to zero, we obtained the optimum valve of $w_{11}$ and $w_{12}$ as

$$w^*_{11} = \frac{1}{1+(1-\rho^2)\theta C_0^2 + \frac{L-1}{n}K\frac{S^2_{Y2}}{\overline{Y}^2}}$$

$$w^*_{12} = \frac{\rho C_0 \overline{Y}}{\overline{X}C_1\left[1+(1-\rho^2)\theta C_0^2\right]+\frac{L-1}{n}K\frac{S^2_{Y2}}{\overline{Y}^2}}$$

$$MSE(t_2) = \overline{Y}^2 + w_{21}^2\overline{Y}^2 A_1 + w_{22}^2\overline{X}^2 A_2 + 2w_{21}w_{22}\overline{X}\overline{Y}A_3 - 2w_{21}\overline{Y}^2 A_4 - 2w_2\overline{X}\overline{Y}A_5 + w_{21}^2 A_6 \quad (2.8)$$

Where $A_1 = 1 + \theta(C_0^2 + (2\delta^2 + \delta)C_1^2 - 4\delta\rho C_0 C_1)$

$A_2 = \theta C_1^2$

$A_3 = \theta(2\delta C_1^2 - \rho C_0 C_1)$

$A_4 = 1 - \theta(\delta\rho C_0 C_1 - \frac{\delta(\delta+1)}{2}C_1^2)$

$A_5 = \delta\theta C_1^2$

$A_6 = \frac{L-1}{n}KS^2_{Y2}$

Partially differentiating the above expression with respect to $w_{21}$ and $w_{22}$, and then equating to zero, we obtained the optimum valve of $w_{21}$ and $w_{22}$ as

$$w^*_{21} = \frac{A_2 A_4 - A_3 A_5}{A_1 A_2 - A_3^2 + \frac{A_2 A_6}{\overline{Y}^2}}$$

$$w^*_{22} = \frac{\overline{Y}}{\overline{X}}\left[\frac{A_1 A_5 - A_3 A_4 + \frac{A_5 A_6}{\overline{Y}^2}}{A_1 A_2 - A_3^2 + \frac{A_2 A_6}{\overline{Y}^2}}\right]$$

$$MSE(t_3) = \gamma^2\left[\overline{Y}^2(1+\theta C_0^2) + b^2\overline{X}^2\theta C_1^2 - 2\overline{X}\overline{Y}b\rho\theta C_0 C_1\right] + \overline{Y}^2(1-2\gamma) + \frac{L-1}{n}K\gamma^2 S^2_{Y2} \quad (2.9)$$

Partially differentiating the above expression with respect to γ, and then equating to zero, we obtained the optimum valve of γ as

$$\gamma^* = \frac{\overline{Y}^2}{\left[\overline{Y}^2(1-\theta C_0^2) + b^2\overline{X}^2\theta C_1^2 - 2\overline{X}\,\overline{Y}b\rho\theta C_0 C_1 + \frac{L-1}{n}KS_{Y2}^2\right]}$$

## 3. Empirical Study

For numerical illustration, we have considered the data given in Murthy (1967, p. 131- 132). The data are based on length (X) and timber volume (Y) for 176 forest strips. Murthy (1967, p. 149) and Kushwaha and Singh (1989) reported the values of intraclass correlation coefficients $\rho_X$ and $\rho_Y$ approximately equal for the systematic sample of size 16 by enumerating all possible systematic samples after arranging the data in ascending order of strip length. The details of population parameters are:

$N = 176$, $\quad n = 16$, $\quad \overline{Y} = 282.6136$, $\quad \overline{X} = 6.9943$,

$S_Y^2 = 24114.6700$, $\quad S_X^2 = 8.7600$, $\quad \rho = 0.8710$,

$S_{Y2}^2 = \frac{3}{4}S_Y^2 = 18086.0025$.

The Table 3.1 shows the percentage relative efficiency (PRE) of proposed estimators with respect to $\overline{y}^{**}$ for the different choices of K and L.

**Table 3.1** : PRE of proposed estimators with respect to usual estimator

| K | L | PRE of $\bar{y}_{lr}^{**}$ with respect to $\bar{y}^{**}$ | PRE of $t_1$ with respect to $\bar{y}^{**}$ | PRE of $t_2$ with respect to $\bar{y}^{**}$ (when $\delta = 1$ and $\alpha = 0$) | PRE of $t_3$ with respect to $\bar{y}^{**}$ (when $\gamma = 1$) |
|---|---|---|---|---|---|
| 0.1 | 2.0 | 407.4884 | 434.0181 | 438.9431 | 434.0199 |
|  | 2.5 | 404.1824 | 430.7801 | 435.7177 | 430.7819 |
|  | 3.0 | 400.9468 | 427.6068 | 432.5561 | 427.6086 |
|  | 3.5 | 397.7794 | 424.4964 | 429.4564 | 424.4982 |
| 0.2 | 2.0 | 400.9468 | 427.6068 | 432.5561 | 427.6086 |
|  | 2.5 | 394.6779 | 421.4470 | 426.4168 | 421.4487 |
|  | 3.0 | 388.6647 | 415.5240 | 420.5112 | 415.5257 |
|  | 3.5 | 382.8921 | 409.8246 | 414.8261 | 409.8262 |
| 0.3 | 2.0 | 394.6779 | 421.4470 | 426.4168 | 421.4487 |
|  | 2.5 | 385.7493 | 412.6472 | 417.6419 | 412.6488 |
|  | 3.0 | 377.3458 | 404.3362 | 409.3494 | 415.5257 |
|  | 3.5 | 369.4225 | 396.4225 | 401.5007 | 409.8262 |
| 0.4 | 2.0 | 388.6647 | 415.5240 | 420.5112 | 421.4487 |
|  | 2.5 | 377.3458 | 404.3362 | 409.3494 | 412.6488 |
|  | 3.0 | 366.8810 | 393.9475 | 398.9770 | 404.3379 |
|  | 3.5 | 357.1773 | 384.2753 | 389.3132 | 369.4760 |

## 5. Conclusion

In this paper, we have proposed A general class of modified regression type estimators for estimating the population mean in systematic sampling using auxiliary information in the presence of non-response. From the above empirical study we see the PRE of all estimators are decreasing with increasing non-response rate K as well as with increasing L. And here we see that in all proposed estimators, $t_2$ gives better result under non-response than other proposed estimators.